\documentclass[prl,twocolumn,showpacs,debugon,nofootinbib]{revtex4}
\usepackage{epsfig}
\usepackage{color}
\newcommand{\bk}{{\bf k}}
\newcommand{\be}{\begin{equation}}
\newcommand{\ee}{\end{equation}}
\renewcommand{\r}{{\bf r}}

\newcommand{\R}{{\bf R}}

\newcommand{\ep}{\epsilon}

\begin{document}


\title{Superradiance and multiple scattering of photons  in atomic gases}
\author{A. Gero and E. Akkermans}
\affiliation{Department of Physics, Technion - Israel Institute of
Technology,
   Haifa 32000, Israel}

\begin{abstract}
We study the influence of cooperative effects such as superradiance and subradiance, on the scattering properties of dilute atomic gases. We show that cooperative effects  lead to an effective potential between pairs of 
atoms that decays like $1/r$. In the case of superradiance, this
potential is attractive for close enough atoms and can be
interpreted as a coherent mesoscopic effect.  We consider a model of multiple scattering of a photon among  superradiant pairs and calculate the elastic mean free path and the group velocity. We study first  the case of a scalar wave which allows to obtain and to understand basic features of cooperative effects and multiple scattering. We then turn to the general problem of a vector wave. In both cases, we obtain qualitatively similar results and  derive, for the case of a scalar wave, analytic expressions of the elastic mean free path and of the group velocity for an arbitrary (near resonance) detuning. 

\end{abstract}


\date{\today}

\maketitle
\section{I. Introduction}
Coherent multiple scattering of photons in cold
atomic gases is an important problem since it presents a path towards the
onset of Anderson localization transition, a long standing and
still open issue.  The large resonant scattering cross-section of
photons reduces the elastic mean free path to values comparable to
the photon wavelength for which the weak disorder approximation
breaks down thus signaling the onset of Anderson localization
transition \cite{am,rama}. Another advantage of cold atomic gases
is that sources of decoherence and inelastic scattering such as
Doppler  broadening are often negligible. Moreover,  propagation of
photons in atomic gases differs  from the case of electrons in
disordered metals or of electromagnetic waves in suspensions of
classical scatterers for which mesoscopic effects and Anderson
localization have been thoroughly investigated \cite{am}. This
problem is then of great interest since it may raise new issues in
the Anderson problem such as change of universality class and
therefore new critical behavior.  New features displayed by the
photon-atom problem  are the existence of  internal degrees of
freedom (Zeeman sublevels) and cooperative effects such as
subradiance or superradiance \cite{dicke} that lead to effective interactions
between atoms \cite{ketterle}. These two differences are expected to lead to
qualitative changes of  both mesoscopic quantities and Anderson
localization. Some of the effects  of a  Zeeman degeneracy have
been investigated in the weak disorder limit \cite{miniat} using a
set of finite phase coherence times \cite{amm} which reduce
mesoscopic effects, such as coherent backscattering
\cite{am,akkmayn}. 

 The influence of cooperative effects and more specifically of
superradiance on the multiple scattering of photons has been recently investigated \cite{GA}.  
It has been shown that in atomic gases  superradiance and subradiance  lead to a  potential between two
atoms, analogous to the one considered in
\cite{thiru, kurisky}, that decays like the inverse of the distance between them. In the case of superradiance, this
potential is attractive for close enough atoms, and can be
interpreted as a coherent mesoscopic effect. The contribution of
superradiant  pairs to multiple scattering properties of a dilute
gas has been calculated by using an effective propagator that describes a scalar wave being scattered by a pair of two-level atoms.
Simple expressions for the photon elastic mean free path and group velocity have been derived at resonance and found to be significantly  different from that of independent atoms. To be more specific, near
resonance, as well as at resonance, the superradiant effect leads to a finite and positive group velocity,
unlike the one obtained for light interaction with independent
atoms.

In this paper we provide, for the case of a scalar wave,  closed expressions  for the suprerradiant contribution to the elastic mean free path and the group velocity for an arbitrary (near resonance) detuning, and calculate the dependence of the transport time on it. In addition, we estimate the maximal inter-atomic separation in a superradiant pair, which accounts for  possible mechanisms that may break the pair. We also compare the effective approach presented in 
\cite{GA} to a more realistic one that takes into account the vectorial nature of the wave.

The paper is organized as follows: In section II we describe the
model which consists of pairs of two-level atoms placed in an external
radiation field where the Doppler shift and recoil effects are
negligible. In order to investigate the influence of the
cooperative effects of such pairs on the multiple scattering of
photons we briefly review, in section III, Dicke states and some of
their properties. Then, we calculate the average interaction
potential of a pair of atoms in a Dicke state by averaging upon
the random orientations of pairs of atoms with respect to the wave
vector of a photon incident on the atomic cloud. Next, we study
the scattering of a photon by such pairs and, in section
IV, compare the results to the case where a classical wave is being
scattered by a pair of atoms. This comparison allows to find an unexpected
connection between superradiance and mesoscopic effects. In sections V
and VI, we consider the multiple scattering of photons by pairs of atoms
and calculate the elastic mean free path
and the group velocity of photons in the random medium. Finally, our analysis is compared to other approaches
in part VII and its results are summarized in part VIII.

\section{II. Model}
Atoms are taken to be degenerate, two-level systems denoted by $|g
\rangle= |j_{g}=0,m_{g}=0 \rangle$ for the ground state and $|e
\rangle = |j_{e}=1,m_{e}=0,\pm1 \rangle$ for  the excited state,
where $j$ is the total angular momentum and $m$ is its  projection
on a quantization axis, taken as the $\hat{z}$ axis. The energy
separation between the two levels, including radiative shift,  is
$\hbar\omega_{0}$ and the natural width of the excited level is
$\hbar \Gamma$. This simple picture of a two-level atom neglects
the rather complicated energy structure of a real atom which reflects
various internal interactions, {\it e.g.}, Coulomb interactions,
spin-orbit interactions, hyperfine interactions, etc. But, due to
selection rules which limit the allowed transitions between
states, in some cases a certain state may couple to only one
other. Thus, the two-level atom approximation is close to reality
and not merely a mathematical convenience.

 We consider a pair of such atoms in an external radiation field and the corresponding
Hamiltonian is $H = H_0 + V$, with \be H_{0}={\hbar\omega_{0}
\over 2} \sum_{l=1}^{2}(|e\rangle\langle e|-|g\rangle\langle
g|)_{l}+ \sum_{\bf{k}\varepsilon} \hbar \omega_{k}
a_{\bf{k}\varepsilon}^{\dag}a_{\bf{k}\varepsilon} \label{eq1}. \ee
$a_{\bf{k}\varepsilon}$ (resp. $a_{\bf{k}\varepsilon}^{\dag}$) is
the annihilation (resp. creation) operator of a mode of the field
of wave vector $\bf{k}$, polarization $\hat \varepsilon_{\bf{k}}$
and angular frequency $\omega_{k}=c|\bf{k}|$.  The interaction $V$
between the radiation field and the dipole moments of the atoms is
given by \be V=-\sum_{l=1}^{2}
\textbf{d}_{l}\cdot\textbf{E}(\textbf{r}_{l})\label{eq2} \ee where
${\bf E}(\r )$ is the electric field operator
\be\textbf{E}(\textbf{r})=i\sum_{\bf{k}\varepsilon}\sqrt{\frac{\hbar\omega_{k}}{2\epsilon_{0}\Omega}}(a_{\bf{k}\varepsilon}\hat
\varepsilon_{\bf{k}}e^{i\bf{k}\cdot
r}-a_{\bf{k}\varepsilon}^{\dag}\hat
\varepsilon_{\bf{k}}^{*}e^{-i\bf{k}\cdot r})\label{eq3}.\ee 
$\Omega\ $ is a quantization volume and
$\textbf{d}_{l}$ is the electric dipole moment operator of the
$l$-th atom. As an odd operator, which changes sign upon inversion, $\textbf{d}_{l}$ may be written as \be
\textbf{d}_{l}=\langle g|\textbf{d}|e\rangle\Delta_{l}^{-}+\langle
e|\textbf{d}|g\rangle\Delta_{l}^{+}\label{eq4} \ee where the
atomic raising and lowering operators are \be\begin{array}{cc}
 \Delta^{+}_{l}=(|e\rangle\langle g|)_{l}  & \Delta^{-}_{l}=(|g\rangle\langle e|)_{l}\end{array}\label{eq5}.\ee

 We assume that the typical speed of the atoms, $v\simeq \sqrt{k_BT_0/ \mu}$, is small compared to
$v_{max}=\Gamma/k$ but large compared to $v_{min}=\hbar k / \mu$ where $\mu$ is
the mass of the atom and $T_0$ is the temperature, so that it is possible to neglect the
Doppler shift and recoil effects. Indeed, for a temperature of
$T_0=10^{-3}K$, the typical speed of the atom is $v\simeq 0.3 m/s$.  Since, for a
wave number of $k=10^{7} m^{-1}$ and
a natural width of $\Gamma=10^{7} s^{-1}$, $v_{max}\simeq 1m/s$ and $v_{min}\simeq0.01m/s$ both assumptions are
fulfilled.

\section{III. Dicke States}

\subsection{A. Interaction potential and lifetime}
The absorption of a photon by a pair of atoms in their ground
state leads to a configuration where the two atoms, one excited
and the second in its ground state, have multiple exchange of a
photon, giving rise to an effective interaction potential and to a
modified lifetime as compared to independent atoms. These two
quantities are obtained from  the matrix elements of the evolution
operator $U(t)$ between states such as $|g_1 e_2 ; 0 \rangle$.
There are six unperturbed and degenerate states with no photon,
given by $\{ |g_{1}  e_{2i} ; 0 \rangle, |e_{1j} g_{2} ; 0 \rangle
\}$  in a standard basis where $i,j=-1,0,1$.  The symmetries of
the Hamiltonian, namely its invariance by rotation around the axis
between the two atoms, and by reflection with respect to a plane
containing this axis, allows to use combinations of these
states that are given by \be|\phi_i ^\epsilon \rangle = {1 \over
\sqrt{2}} [ |e_{1i} g_2 ; 0 \rangle + \epsilon |g_{1} e_{2i} ; 0
\rangle ] \label{eq6}\ee with $\epsilon = \pm 1$, so that
\be\langle \phi_j ^{\epsilon'} | U(t) | \phi_i ^\epsilon \rangle =
\delta_{ij} \delta_{\epsilon \epsilon'} S_i ^\epsilon (t)
\label{eq7}\ee and \be S_i ^\epsilon (t) = \langle e_{1i} g_2 ; 0
| U(t) | e_{1i} g_2 ; 0 \rangle + \epsilon \langle g_{1} e_{2i} ;
0 | U(t) | e_{1i} g_2 ; 0 \rangle \label{eq8}. \ee

 The states $|\phi_i ^\epsilon \rangle$ may be rewritten in terms of the
well-known Dicke states $|L M \rangle$, where $L$ is the
cooperation number and $M$ is half of the total atomic inversion
\cite{dicke}. For two atoms, the singlet Dicke state is \be|00
\rangle = {1 \over \sqrt{2}} [ |e_{1} g_2 \rangle - |g_{1} e_{2}
\rangle ] \label{eq8a}\ee and the triplet Dicke states are 
\begin{eqnarray}
|11 \rangle &=& |e_{1} e_2  \rangle  \nonumber \\ |10 \rangle &=& {1 \over \sqrt{2}} [ |e_{1} g_2 \rangle + |g_{1} e_{2} \rangle] \nonumber  \\ |1-1\rangle &=&|g_{1} g_2 \rangle
\label{eq8b}
\end{eqnarray}
The states $|11 \rangle$ and $|1-1 \rangle$ correspond,
respectively, to both atoms in their excited states and both atoms
in their ground state. The singlet state $|00 \rangle$ and the
triplet state $|10 \rangle$ both correspond to one atom in the
excited state and the other in the ground state, but $|00 \rangle$
is anti-symmetric  where $|10 \rangle$ is symmetric under an
exchange of the atoms. Therefore, we may rewrite (\ref{eq6}) as
 $ |\phi_i ^+ \rangle = |1 0 ; 0 \rangle $ and
$|\phi_i ^- \rangle = |0 0 ; 0\rangle$.

 For times such that $t \gg
r/c$, where $r$ is the distance between the two atoms,  up to
second order in the coupling to the radiation, (\ref{eq8}) reads \be
S_i ^\epsilon (t) \simeq 1 - {it \over \hbar} \left( \Delta E_i
^\epsilon - i {\hbar \Gamma_i ^\epsilon \over 2}\right ) . \label{eq9}
\ee The two real quantities $\Delta E_i ^\epsilon $ and $\Gamma_i
^\epsilon $ are respectively the interacting potential  and the
probability per unit time of emission of a photon by  the two
atoms in the state $|\phi_i ^\epsilon \rangle$. The calculation of
these two quantities requires second order perturbation theory
with respect to the interaction (\ref{eq2}). To this purpose we
define an initial state where one atom is excited and the other is
in its ground state without any photon and a final state where the
two atoms are exchanged. We also define intermediate states of
two types: both  atoms in their ground state with one virtual
photon present and both atoms in their excited state with one virtual
photon present. Summing the corresponding  diagrams
\cite{stephen} gives  \be \Delta E_i ^\ep = \ep {3 \hbar \Gamma
\over 4} \left[ - p_i {\cos k_0 r \over k_0 r} + q_i \left( { \cos
k_0 r \over (k_0 r)^3 } + {\sin k_0 r \over (k_0 r)^2} \right)
\right] \label{eq10} \ee and \be {\Gamma_i ^\epsilon \over \Gamma}
= 1 - {3 \over 2} \ep \left[ - p_i {\sin k_0 r \over k_0 r} + q_i
\left( {\sin k_0 r \over (k_0 r)^3} -  { \cos k_0 r \over (k_0
r)^2 } \right) \right] \label{eq11}\ee where we have defined
$k_0=\omega_0/c$,\be
p_i = 1 - {\hat \r}^2_i  \, ,   q_i = 1 - 3 \,  {\hat \r}^2_i
\label{eq12}\ee  and $\hat{\r} =(1,\theta,\varphi)$
being a unit vector along the direction joining the two atoms. For a $\Delta m=m_e-m_g=0$ transition, \be
p_{0} = \sin^{2} \theta \,  ,     q_{0} = 1 - 3 \cos^2\theta
\label{eq13}\ee while for a $\Delta m=\pm1$ transition, \be
p_{\pm} = \frac{1}{2}(1+\cos^2\theta)    \,  ,
q_{\pm} = \frac{1}{2}(3\cos^2\theta-1).
\label{eq14}\ee
 At short distance $k_0 r
\ll 1$, we obtain that $\Gamma_i ^+ \simeq 2 \Gamma$ for the
superradiant state $|\phi_i ^+ \rangle = |1 0 ; 0\rangle$ and
$\Gamma_i ^- \simeq 0 $ for the subradiant state $|\phi_i ^- \rangle =
|0 0 ; 0\rangle$.

\subsection{B. Average interaction potential}
 For a photon of wave vector $\bk$ incident on an atomic cloud, the
potential between  two atoms that we shall denote by $V_e$ is obtained from (\ref{eq10}) by averaging upon
the random orientations of the pairs of atoms with respect to
$\bk$. Since, according to (\ref{eq13}) and (\ref{eq14}), $\langle
q_i \rangle =0$ and $\langle p_i \rangle = 2/3$,
we obtain for the average potential $V_{e}$ \be \ep V_{e} (r) =
\langle \Delta E_i ^\ep \rangle  = - \ep {\hbar \Gamma \over 2}
{\cos k_0 r \over k_0 r} \label{eq15} \ee and the average inverse
lifetimes of Dicke states are \be\langle \Gamma_i ^\epsilon
\rangle= \Gamma \left( 1 + \ep {\sin k_0 r \over k_0 r} \right)
\label{eq16} \ee which retain the same features as (\ref{eq11})
for $k_0 r \ll 1$.

\medskip

Let us now characterize the interaction potential $V_e$.  Whereas for
a single pair of atoms,  the potential (\ref{eq10})  is
anisotropic and decays at short distance like $1/ r^3$, a behavior
that originates from the transverse part of the photon propagator,
we obtain that on average over angular configurations, the
potential (\ref{eq15}) between two atoms in a Dicke state $|L 0
\rangle$ in vacuum becomes isotropic and decays like $1/ r$. This
behavior coincides with the one obtained by considering the interaction of
two-level atoms with a scalar wave.  This  could have been
anticipated since in that case the transverse contribution $q_i$ to the photon
propagator averages to 0.  A related  behavior for the
orientation average interaction potential has been also obtained for
the case of  an intense radiation field \cite{thiru} and it has
recently been investigated in order to study effects of a long
range and attractive potential between atoms in a Bose-Einstein
condensate for a  far detuned light \cite{kurisky}. This latter
potential, which is fourth order in the coupling to the radiation,
 corresponds to the interaction energy between two atoms in their
ground state in the presence of at least one photon. The average
potential $V_e$ we have obtained is different from that case: it is
second order  in the coupling to the radiation and it corresponds to
the interaction energy of Dicke states $|L 0 \rangle$ in vacuum.

\subsection{C. Scattering properties}
In order to study the scattering properties of Dicke states we
introduce the collision operator $T(z)=V+VG(z)V$ where $V$ is
given by (\ref{eq2}) and $G(z)=(z-H)^{-1}$ is the resolvent where
the Hamiltonian $H$ is the sum of (\ref{eq1}) and (\ref{eq2}). The
matrix element that describes the transition amplitude from the initial
state $|i \rangle = |1-1; {\bk} {\hat \varepsilon} \rangle$, where
the two atoms  are in their ground state in the presence of a
photon of frequency $\omega=c|\bk|$ and polarization $\hat \varepsilon$, to the final state $|f \rangle
= | 1-1; {\bk}'{\hat \varepsilon}' \rangle$ is \be T=\langle f|
T(z=\hbar(\omega - \omega_0))|i\rangle \label{eq17}\ee where $|\bf{k}|=|\bf{k'}|$. By using
the closure relation we may write $T$ as the sum of a superradiant
and a subradiant contribution, $T = T^{+} + T^{-}$ \cite{rk}
with \be T^{\pm} = \langle f | V | \phi^{\pm} \rangle \langle
\phi^{\pm}| G (z=\hbar( \omega - \omega_0)) | \phi^{\pm} \rangle
\langle \phi^{\pm} | V | i \rangle \label{eq18} \ee 
where $|\phi^{\pm}>$ are the Dicke states $|L0>$ in vacuum. The two matrix
elements in (\ref{eq18})  represent the absorption and the
emission of a real photon by the pair of atoms. They are easily obtained
from (\ref{eq2})-(\ref{eq5}) and lead to the following expressions
for the scattering amplitudes \be T ^{+} = A e^{ i (\bk - \bk') \cdot \R}
\cos \left( {\bk \cdot \r \over 2} \right)\cos \left( {\bk' \cdot
\r \over 2} \right) G ^{+} \label{eq19} \ee and \be T ^{-} = A e^{
i (\bk - \bk') \cdot \R } \sin \left( {\bk \cdot \r \over 2}
\right)\sin \left( {\bk' \cdot \r \over 2} \right) G ^{-}
\label{eq20} .\ee We have defined $\r = \r_1 - \r_2$ , ${\bf R} =
({\bf r_1} + {\bf r_2}) / 2$ and \be A = {\hbar \omega \over \ep_0
\Omega} d ^2 ({\hat d} \cdot {\hat \varepsilon} ) ({\hat d}^*
\cdot {\hat \varepsilon}'^* )\label{eq21}\ee where the reduced
matrix element and the corresponding unit vector are
 \be 
d=\frac{\langle j_e\|\textbf{d}\|j_g\rangle}{\sqrt{2j_e+1}} \,   \, ,
 \hat{d}=\frac{1}{d}\langle j_em_e|\textbf{d}|j_gm_g\rangle \, .
\label{eq21a}
\ee
 The propagators $G^{\pm}$ are the
expectation values of the resolvent in the Dicke states
$|\phi^{\pm}\rangle$, namely $ G ^{\pm} = \langle \phi^{\pm}| G(
\hbar \delta) | \phi^{\pm}\rangle$ where close to resonance
$\delta = \omega-\omega_{0} \ll \omega_{0}$. The propagators result from the
sum of an infinite series of virtual photon exchanges between the
two atoms in the pair and are given in  terms of (\ref{eq10}) and
(\ref{eq11}) by \be G ^{\pm} = {\left( \hbar \delta -\Delta
E^{\pm}+i  {\hbar} \frac{\Gamma^{\pm}} {2}\right)^{-1}} \label{eq22} . \ee

The average propagator is then obtained by averaging 
$ G ^{\pm}$  over the random orientations of the pairs of atoms with respect to the wave vector $\bk$ of the incident photon. However, we shall consider in a first stage the effective propagator obtained for the case of a scalar wave. This accounts to write for the effective propagator the expression
\be G_{e} ^{\pm} = {\left [ \hbar \left(\delta + i{\Gamma \over 2} \pm {\Gamma \over 2} {e^{i k_0 r} \over k_0 r}\right)\right]^{-1} }
\label{eq23} \ee where we have used (\ref{eq15}) and (\ref{eq16})
for the average potential and for the average inverse lifetimes.
This expression   constitutes  a priori a rough approximation of the exact average. We shall calculate later, in section VI,  the exact expression of the average propagator and show that it is rather complicated whereas the approximate expression using a scalar wave gives similar  qualitative results. Therefore, it allows  for a better understanding of relevant physical quantities such as elastic mean free path and group velocity. From now on, we thus use the scalar wave approximation in order to provide, in a rather simple way, the main features of multiple scattering by superradiant pairs.

 
 With the help of (\ref{eq23}), the scattering
amplitudes are \be T_{e} ^{+} = A e^{ i (\bk - \bk') \cdot \R}
\cos \left( {\bk \cdot \r \over 2} \right)\cos \left( {\bk' \cdot
\r \over 2} \right) G_{e} ^{+} \label{eq24} \ee and \be T_{e} ^{-}
= A e^{ i (\bk - \bk') \cdot \R } \sin \left( {\bk \cdot \r \over
2} \right)\sin \left( {\bk' \cdot \r \over 2} \right) G_{e} ^{-}
\label{eq25} .\ee At short distances $k_0 r \ll 1$, the subradiant
amplitude $T_{e} ^{-}$ becomes negligible as compared to the
superradiant term $T_{e} ^{+}$. Therefore, the potential
(\ref{eq15}) is attractive and decays like $1/r$.  More
precisely, at short distances the effective propagator $G_{e}^{-}$
diverges for $ \delta /\Gamma = 1/(2 k_0 r)$ and $G_{e}^{+}$ is
purely imaginary for $ \delta / \Gamma = -1/(2 k_0 r)$. Thus, for
$ \delta / \Gamma < 1/(2 k_0 r)$ the imaginary part of the
subradiative term (\ref{eq25}) is negligible as compared to the
imaginary part of the superradiative term (\ref{eq24}) and for
$|\delta|/ \Gamma < 1/(2 k_0 r)$ both the real part and the
imaginary parts of (\ref{eq25}) are negligible as compared to
(\ref{eq24}).

 We can interpret these results by saying that,
at short distances $(k_0 r \ll 1)$, the time evolution of the
initial state \be| \psi (0) \rangle = |e_1g_2; 0 \rangle = {1
\over \sqrt{2}} [ |\phi^{+} \rangle + |\phi^{-} \rangle
]\label{eq26}\ee corresponds, for times shorter than $1 / \Gamma$,
to a periodic exchange of a virtual photon between the two atoms
at the Rabi frequency \be \Omega_{R}=\frac{\langle \Delta E^- \rangle -
\langle \Delta E^+ \rangle}{\hbar} \label{eq27}\ee which is much larger than $ \Gamma$ since, with the help of  (\ref{eq15}), \be \Omega_{R} \simeq \frac{\Gamma}{k_0
r}.\label{eq27a}\ee For larger
times,  the two atoms return to their ground state and a real
photon $(\bk' {\hat \varepsilon}')$ is emitted. At large distances
$(k_0 r \gg 1)$, the Rabi frequency becomes smaller than $\Gamma$,
so that the excitation energy makes only a few oscillations
between the two atoms, thus leading to a negligible interaction
potential.

 We  finally notice that the angular
distribution of the light scattered by two atoms in a superradiant
state is nearly identical to that of a single atom. This follows from the fact that at short
distance $k_0 r \ll 1$, we can neglect higher order multipolar corrections so that the corresponding additional phase shift, $k_0 r
\cos \vartheta$, between waves emitted by the two atoms becomes negligible ($\vartheta$ is the angle between
the direction of the emitted photon and the axis between the two
atoms).

\section{IV. Cooperative Effects and coherent backscattering}

\begin{figure}[ht]
\centerline{ \epsfxsize 4cm \epsffile{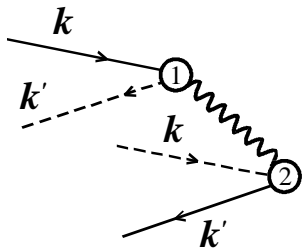} } \caption{\em
Schematic representation of the two amplitudes that describe
double scattering of a scalar wave.
The wavy line accounts for the exchange of a virtual photon between the two
atoms. This diagram is analogous to the coherent backscattering in quantum mesoscopic physics. } \label{fig1}
\end{figure}

It is interesting to derive the previous results in another way
that emphasizes the analogy with coherent backscattering
\cite{am,rama}. To that purpose, we write the scattering amplitude
$T$ defined previously in  (\ref{eq17}) as a superposition of  two ''classical'',
scalar amplitudes, $T_1$ and $T_2$ \cite{rk2},  each of them being
a sum of single scattering and double scattering contributions,
that is  \be T_1 = {t \over 1 - t^2 G_0 ^2} \left[ e^{i (\bk -
\bk') \cdot \r_1} + t \, G_0 \, e^{i (\bk \cdot \r_1 - \bk' \cdot \r_2)}
\right] \label{eq28} \ee and \be T_2 = {t \over 1 - t^2 G_0 ^2}
\left[ e^{i (\bk - \bk') \cdot \r_2} + t \, G_0 \, e^{i (\bk \cdot \r_2
- \bk' \cdot \r_1)} \right]  \label{eq29}. \ee Here \be t
=\frac{4\pi}{k_0}\frac{\frac{\Gamma}{2}}{\delta +
i\frac{\Gamma}{2}}\label{eq30} \ee is the amplitude of a scalar
wave scattered by a single atom at the origin and the prefactor $t
/ ( 1 - t^2 G_0 ^2)$ where \be G_0 =-\frac{e^{ i k_0 r}}{4 \pi
r}\label{eq31} \ee accounts for the summation of the series of
virtual photon exchange between the two scatterers.  We can single
out  in the total amplitude $T= T_1 + T_2$  the single scattering
contribution $T_{s}$ and write the intensity associated to the
higher order scattering term shown in Figure \ref{fig1} as \be |T -
T_{s}|^2 = 2 \left| {t^2 G_0 \over 1 - t^2 G_0 ^2 } \right|^2
\left[ 1 + \cos (\bk + \bk' ) \cdot (\r_1 - \r_2 ) \right]
\label{eq32} .\ee

The structure of relation (\ref{eq32}) is very reminiscent  to that of the so-called coherent backscattering intensity  which occurs in the multiple elastic scattering of light.
But although they are analogous, (\ref{eq32}) differs from coherent backscattering. In the latter case,  averaging over the spatial positions $\r_1$ and $\r_2$ makes the interference term $\cos (\bk + \bk' ) \cdot (\r_1 - \r_2 )$ vanish in general, with two exceptions:

$\bk+\bk'\simeq0$ : In the direction exactly opposite to the direction of incidence, the intensity is twice the classical value. This phenomenon is known as coherent backscattering.

$\r_1=\r_2$ : Closed multiple scattering trajectories  which are at the origin of the phenomenon of weak localization. 

 In (\ref{eq32}) the interference term, {\it i.e.,} the second term in the bracket,
reaches its maximum value 1 for $\r_1 = \r_2$
so that we obtain from (\ref{eq28})-(\ref{eq29}) and (\ref{eq24})
that $T_1 = T_2 \propto (1/2) T_{e} ^+$, up to a proportionality
factor \cite{rk2}. Thus, the total amplitude is  given by
the superradiant term with no subradiant contribution.

\section{V. Multiple Scattering and Cooperative Effects}
\subsection{A. Effective self-energy}
 We consider now multiple scattering of a photon by superradiant pairs built out of atoms separated by a distance $r$ and coupled by the attractive interaction potential $V_{e}$.
 This situation corresponds to a dilute gas that is assumed to fulfill \be r \ll \lambda_0 \ll n_i ^{-1/3}\label{32aa}\ee where $n_i$ is the density of pairs and $\lambda_0 = 2 \pi / k_0$ is the atomic transition wavelength.
  The limiting case (\ref{32aa}) corresponds to a situation where the two atoms that form  a superradiant pair, through exchange of a virtual photon, constitute an effective scatterer and  cooperative interactions between otherwise well-separated pairs are negligible. Let us stress that  we study here a simplified model where only pairs of atoms have been taken into account. A more realistic model should include higher order terms that account for cooperative effects between more than two atoms, but we do not consider such higher order terms, {\it i.e.,} including superradiant clusters of three or more atoms.
The purpose of the current model is to examine the $\it{contribution}$ of superradiant pairs to the transport properties of the gas. We use the Edwards model \cite{am,cargese} to 
describe the medium as a discrete collection of $N_i$
superradiant pairs in a volume $\Omega$. Each pair, located at 
$\R_l$, is characterized by its scattering potential
$u(\R-\R_l)$.
 Therefore, the  disorder
potential is given by \be U(\R)=\sum_{l=1}^{N_i}u(\R-\R_l)\label{32b}.\ee
We assume that the scattering potential is short range compared to the wavelength, and we approximate it by a (conveniently regularized) delta function potential, $u(\R)=u_0\delta(\R)$. In the limit of a high density of weakly scattering pairs, but with a constant value of $n_iu_{0}^2$, it can be shown
\cite{am} that the correlation function defined by \be B(\R-\R')=n_i\int d\R''u(\R''-\R)u(\R''-\R')\label{32c},\ee becomes 
\be B(\R-\R')=n_iu_0^2\delta(\R-\R')\label{32d}.\ee
 In other words, in this limit, the Edwards model reduces to a Gaussian white noise model characterized by the condition (\ref{32d}).
 
 The Green's function $g$ of a scattered photon is related to the free photon Green's function $g_0$, {\it i.e.}, in the absence of disorder potential, by the equation  \cite{am} \be g=g_0+g_0Ug  \label{32e}.\ee
 Averaging (\ref{32e}) over disorder and using the properties of the Gaussian model discussed above yields the Dyson equation
 \be \langle g \rangle _d=g_0+g_0\Sigma\langle g \rangle_d \label{32f}\ee
 where $\langle \cdot \cdot
\cdot\rangle_d$ denotes averaging over the random potential.
 The function $\Sigma$, known as the self-energy, represents
the sum of all irreducible scattering diagrams. The pertubative
expansion of the self-energy in a power series controlled by the parameter $n_{i}u_{0}^{2}$ is represented 
in Figure 2.

\begin{figure}[ht]
\centerline{ \epsfxsize 8cm \epsffile{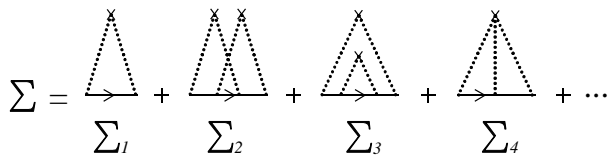} } \caption{\em
Pertubative expansion of the self-energy in a power series in the
parameter $n_iu_{0}^2$. Solid lines account for the free photon Green's function $g_0$. Pairs of dotted lines, connected by $\times$, stand for the two-point correlation function $B$. The first term $\Sigma_{1}$,
proportional to $n_iu_{0}^2$, accounts for independent scattering
events while the second term $\Sigma_{2}$, proportional to
$n_i^2u_{0}^4$, describes interference effects between pairs of 
scatterers. } \label{fig2}
\end{figure}

For small values of $n_iu_0^2$, the main contribution is obtained by
keeping only the first term $\Sigma_{1}$ which describes
independent scattering events. Therefore, the first contribution to the self-energy is proportional to the density of
 scatterers and to the average scattering amplitude and it is given, for  $k_0 r \ll 1$, by
\be \Sigma_{1} =  {6 \pi n_i\over k_0}A_{j_gj_e} \hbar \Gamma
 {\overline G}_e ^+ \label{eq33}\ee where \be
A_{j_gj_e}=\frac{1}{3}\frac{2j_e+1}{2j_g+1}\label{eq32b}.\ee
The latter quantity is obtained by averaging $A$ in  (\ref{eq24}) over Zeeman sublevels $m$ that appear in its definition given by  (\ref{eq21}) and (\ref{eq21a}).

The additional average, denoted by $\overline{\cdot
\cdot \cdot}$, is  taken over distances $r$ up to a maximal value
$r_m $ which accounts for all possible mechanisms
that may break  the pairs. 

The value of $r_m$ can be estimated by comparing the kinetic energy $K$ of a superradiant pair  to its  average potential energy $V_e^+$. We have 
 $K\simeq \hbar^2 / \mu r^2$ and from (\ref{eq15}) we obtain  that $V_e^+ \simeq -\hbar\Gamma/2k_0r$.
 Minimizing the average energy \be E(r)\simeq\frac{\hbar^2}{\mu r^2} -\frac{\hbar\Gamma}{2k_0r}\ee with respect to $r$ yields \be k_0 r_m=4\frac{\hbar k_0^2}{\mu\Gamma}\label{eq32c}\ee or  \be k_0 r_m=4\frac{v_{min}}{v_{max}}\label{eq32d} \ee where the speeds $v_{min}$ and $v_{max}$ have been defined in section II.
 For typical values, ${\Gamma } = 10^7s^{-1}$ and $k_{0}=
10^{7}m^{-1}$ we obtain that $k_0r_m\simeq 0.05$.
 Thus, we can use the results obtained in subsection III-C and consider the superradiant term only.

For $j_g=0$ and $j_e=1$, $A_{01}=1$  and using (\ref{eq23}) we rewrite (\ref{eq33}) as \be \Sigma_1 =  {
6 \pi n_i \over k_0 } \frac{1}{r_m}\int_0^{r_m} {dr \over { \delta \over
\Gamma} + {1 \over 2k_0 r} + i} \label{eq34} .\ee We stress again that, in our approach, a pair of atoms in a superradiant
state is considered as a single scatterer and  the effective medium
parameters are derived from $\Sigma_{1}$ as it
will be shown in the next subsections. In contrast to our
treatment, others \cite{bart3,morice} consider multiple scattering
of a real photon by \textit{independent} atoms and use the second term
$\Sigma_{2}$, which describes interference effects between the
scatterers, to calculate corrections to the elastic mean
free path and to the refractive index of the medium. A further comparison
between these two points of view is given in section VII.

\subsection{B. Elastic mean free path}
The elastic mean free path $l_e$ is obtained from the imaginary part of
the self-energy, namely
 \be \frac{k_0 }{l_e} = - \mbox{Im} \Sigma_1\label{eq35}.\ee
 Let us stress that (\ref{eq35}) is equivalent, in the case of a dilute gas, to the
 known formula \be l_e=\frac{1}{n_i\sigma_e}\label{eq35a}\ee
 where the total cross section, $\sigma_e$, is obtained for $k_0 r \ll 1$
 from (\ref{eq24}) by means of the optical theorem
 \be \sigma_e=-\frac{2\Omega}{\hbar c}
 \mbox{Im} \langle \, \overline{T}_e^+(\textbf{k}=\textbf{k}',\hat \varepsilon=\hat
\varepsilon')\rangle_m\label{eq35b}\ee and $\langle \cdot \cdot
\cdot\rangle_m$ represents an averaging over Zeeman sublevels. The
equivalence in this case is proven easily if one uses (\ref{eq32b}) and the
usual expression for the inverse lifetime \be
\Gamma=\frac{d^2k_0^3}{3\pi\epsilon_0\hbar}\label{eq35c}\ee where
the reduced matrix element is defined in (\ref{eq21a}). Therefore,
from (\ref{eq34}) and (\ref{eq35}) we obtain that
 \be{1 \over l_e(\delta)} = {6 \pi n_i \over k_0 ^2}f_{1}\left(k_0 r_m, {
\delta \over \Gamma}\right)\label{eq36} \ee where we have defined the
function \be f_{1}(u, v) =\frac{1}{2u} \int_{0}^{2u} {dx \over 1+ 
(v+{1 \over x})^2}
\label{eq37} .\ee
 The integral is easily carried out analytically and the explicit
expression is given in Appendix A. It is interesting to compare
$l_e$ to the elastic mean free path $l_0$ that corresponds to near
resonant elastic scattering of a photon by a single atom. The latter
quantity is obtained by replacing $\Gamma$ by $\Gamma/2$ in
(\ref{eq36}) (since the inverse lifetime of a single atom is  half
 the one related to a superradiant pair) and $1/x$ by $0$ in
(\ref{eq37}) (since the inter-atomic distance is taken to be  infinite for a
single atom) and it is given by \be l_0(\delta) = { k_0 ^2 \over 6
\pi n_i} \left[1 + \left(\frac{2\delta}{\Gamma}\right)^2 \right]
\label{eq38} .\ee
 In Figure 3 the ratio between
these two quantities is plotted as a function of the reduced detuning $\delta/\Gamma$ from resonance for several values of $k_{0}r_{m}$.

\begin{figure}[ht]
\centerline{ \epsfxsize 6cm \epsffile{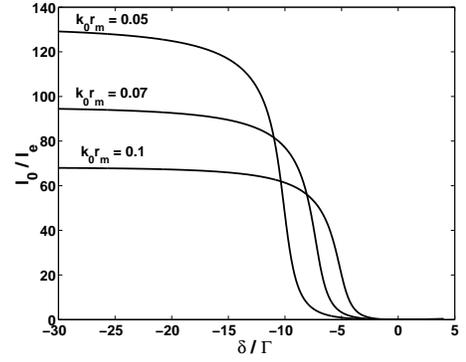} } \caption{\em 
Ratio between the elastic mean free paths $l_{0}$ and  $l_{e}$ as a function of the reduced detuning $\delta/\Gamma$ for 
$k_{0}r_{m}=0.05, 0.07,$ and $0.1$. Away from resonance, for blue
detuning, the elastic mean free path $l_e$ becomes smaller than
$l_0$ in a ratio roughly given by $ 1/(k_0 r_m)^{2}$. At resonance, the ratio between the elastic mean free paths is given by (\ref{eq40}).} \label{fig3}
\end{figure}

 At resonance, we obtain from (\ref{eq36}) that \be
l_{e}(0)=\frac{k_{0}^{2}}{8\pi n_{i}}\frac{1}{(k_{0}r_{m})^2}
\label{eq39}\ee and hence \be{ l_0 (0) \over l_e(0)} = {4 \over 3}
(k_0 r_m )^2 \ll 1 \label{eq40}.\ee Away from resonance, for blue
detuning, the elastic mean free path $l_e$ becomes smaller than
$l_0$ in a ratio roughly given by $ 1/(k_0 r_m)^{2}$. This is a
direct consequence of the existence of the attractive potential
$V_e$.

\subsection{C. Group velocity}
Another  important physical quantity that characterizes multiple
scattering of a photon is its group velocity $v_g$ given in terms
of the refraction index $\eta$ by the usual relation \be \frac{c}{v_g}= \eta + \omega
{d\eta \over d\omega} \label{eq41}.\ee The refraction index
for a dilute medium is
\be\eta=(1+n_i\mbox{Re} \,  \alpha)^{1/2}\label{eq41a} \ee where the
dynamic atomic  polarizability $\alpha$ is proportional to the
self-energy \be \alpha=-\frac{1}{n_i}\left (\frac{c}{\omega}\right
)^2{\Sigma}_1 \label{eq41b}. \ee Thus, we obtain that \be \eta =
\left[ 1 -\left (\frac{c}{\omega}\right )^2 \mbox{Re} \Sigma_1
\right]^{1/2} \label{eq42} .\ee Substituting (\ref{eq42}) into
(\ref{eq41}) yields \be \frac{c}{v_g}= \frac{1}{\eta}\left( 1
-\frac{c^2}{2\omega}\frac{d}{dw}\mbox{Re} \Sigma_1\right)
\label{eq42a} .\ee From the self-energy (\ref{eq34}), we notice
that $\eta \simeq 1$ for all values of the detuning $ \delta /
\Gamma$ and in a large
range of densities  $n_i$ 
 so that \be {c \over v_g(\delta)} \simeq 1-
{n_i \over n_c} f_{2}\left(k_0 r_m, { \delta \over \Gamma}\right)
\label{eq43} \ee where we have defined the characteristic density
\be n_c = {k_0 ^3 \over 6 \pi} {\Gamma \over  \omega_0}
\label{eq44}\ee and the function \be f_{2}(u, v) =\frac{1}{2u}
\int_{0}^{2u} dx {1 - (v + {1 \over x})^2 \over \left(1 + (v + {1
\over x})^2\right)^2} \, .
\label{eq45} \ee The integration is 
easily performed and the explicit expression is given in Appendix
A. By replacing $\Gamma$ by $\Gamma/2$ in (\ref{eq43}) and $1/x$
by $0$ in (\ref{eq45}), we obtain the group velocity  $v_0$ of
light interacting with independent two level-atoms \be{c \over
v_0(\delta)} = 1- {n_i \over n_c}{1 - (\frac{2\delta}{\Gamma})^2
\over \left(1 + (\frac{2\delta}{\Gamma})^2\right)^2} \, .
\label{eq46}
\ee  
For the following typical values,  ${\Gamma } = 10^7s^{-1}$, $k_{0}=
10^{7}m^{-1}$ and $n_i=10^{10}cm^{-3}$, we obtain that $n_i/n_c\simeq 10^5$.
 
Figure 4 displays the group velocities $v_g$ and  $v_0$ plotted as a function of the reduced detuning $\delta/\Gamma$ for ${n_i /
n_c}=10^5$ and $k_{0}r_{m}=0.1$.

\begin{figure}[ht]
\centerline{ \epsfxsize 6cm \epsffile{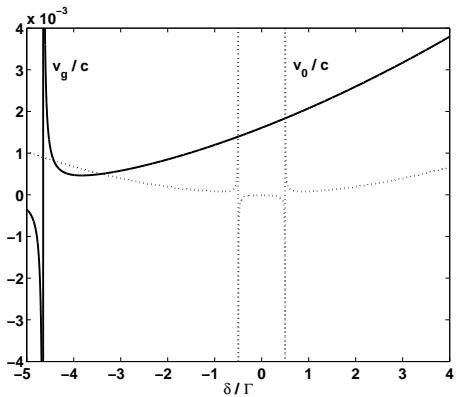} } \caption{\em 
Group velocities $v_{g}$ (solid line) and $v_{0}$ (dotted line) as a function of the reduced 
detuning $\delta/\Gamma$ for ${n_i / n_c}=10^5$ and $k_{0}r_{m}=0.1$. The group velocity
$v_{0}$ diverges at  two symmetric values of order unity of the reduced detuning and it takes negative values in between. The group velocity
$v_{g}$, near resonance, remains finite and positive.}
\label{fig4}
\end{figure}

 $v_g$ appears to diverge at quite a large and negative value
of the detuning $ \delta / \Gamma \simeq -1/(2 k_0 r_m)$.
  But near resonance it is
well behaved, meaning that it remains finite and positive.
At resonance, according to
(\ref{eq43}), the group velocity is \be {c \over v_g (0)} = 1 +4\pi {
n_i \over k_0 ^3}{\omega_0 \over \Gamma} (k_0 r_m)^2
\label{eq47}. \ee

This expression of $v_g$ differs substantially from the one
obtained for $v_{0}$. For densities $n_i > n_c$, the group velocity
$v_{0}$ diverges at  two symmetric values of order unity of the detuning and it takes negative values in between ({\it i.e.}, also at
resonance), as it can be seen in Figure 4. 
This problem has been recognized a long time ago \cite{brillouin}
and an energy velocity has been
defined which describes  energy transport through a diffusive
medium \cite{bart3,loudon}. However, the diffusion coefficient, which
will be discussed in the next subsection, is derived from the
group velocity and not from the energy velocity \cite{am}. 
Moreover, a closed expression for the energy velocity, $v_E$, has been obtained only for the case of a resonant Mie scattering \cite{vt}. The expression is similar to (\ref{eq47}) and is given by \be {c \over v_E} = 1 +9\pi {
n_i \over k_0 ^3}{\omega_0 \over \Gamma} 
\label{eq47a}. \ee
It is then interesting to notice that the inclusion of cooperative effects even at the lowest order, $\it{i.e.}$, taking into account superradiant pairs, allows to obtain a group velocity which is well behaved at resonance, unlike the case of resonant scattering by independent atoms.

\subsection{D. Diffusion coefficient and transport time}

Diffusive transport of photons through a  gas is characterized by
the photon diffusion coefficient \be D(\delta) = {1 \over 3} v_g(\delta)
l_e(\delta) \label{eq48}\ee that combines the elastic mean free
path and the group velocity, both derived from the complex valued
self-energy (\ref{eq34}). The diffusion coefficient $D$ is of
great importance since it enters into expressions of various measured
physical quantities, such as  the  transmission
and the reflection coefficients of a disordered medium \cite{am}. In addition to these average quantities, an incident pulse that probes a nearly static  configuration of scatterers  may provide an instantaneous picture of the medium that displays a random distribution of bright and dark spots. This snapshot, known as a speckle pattern, can be characterized by the  angular-correlation function and the time-correlation function of the light intensity (diffusing wave spectroscopy). In the first case, the correlation function of the transmission coefficient between two distinct directions of the transmitted wave is measured. In the second case, the intensity of the transmitted wave is measured at different times, so that the motion of the scatterers must be taken into account. 
As pointed before, in both cases the diffusion coefficient plays an important role, as it enters in the  relevant expressions. Moreover, the
critical behavior of transport close to the Anderson localization
transition at strong disorder is also obtained from the scaling
form of $D$. Its expression, deduced from (\ref{eq36}) and
(\ref{eq43}), depends on the range $r_m$ and on the detuning $
\delta / \Gamma$.  Since the group velocity and the elastic mean
free path are significantly modified for superradiant states, we
thus expect the diffusion coefficient to  be different from its
value obtained for independent atoms.

We define the transport time by \be \tau_{tr} (\delta
)=\frac{l_e(\delta)}{ v_g(\delta) }
\label{eq49}.\ee At resonance and for $n_i \gg n_c$, it  can be
rewritten with the help of (\ref{eq39}) and (\ref{eq47}) as \be
\tau_{tr} (0) = {1 \over 2 \Gamma} \label{eq50}\ee in accordance with our assumption of superradiant states. Near resonance, the transport time depends weakly on the detuning. But, away from
it, $\tau_{tr}$ depends on the detuning and thus on
 frequency, as it can seen from Figure 5 where the inverse of the transport time $\tau_{tr}^{-1}/\Gamma$ is plotted as a function of the reduced detuning $\delta/\Gamma$ for $n_i= 10^{10}
cm^{-3}$, ${\Gamma} = 10^7s^{-1}$ and $k_{0}=
10^{7}m^{-1}$ for several values of $k_{0}r_{m}$.

\begin{figure}[ht]
\centerline{ \epsfxsize 6cm \epsffile{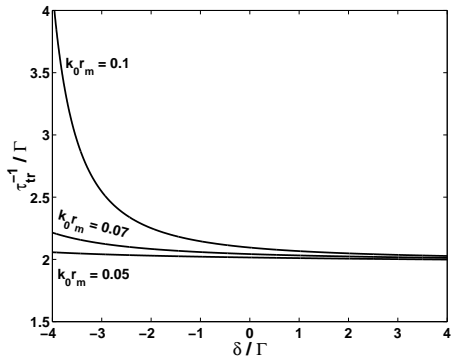} } \caption{\em 
Inverse of the transport time $\tau_{tr}^{-1}/\Gamma$ as a function of  the reduced detuning $\delta/\Gamma$ for $n_i= 
10^{10} cm^{-3}$, ${\Gamma } = 10^7s^{-1}$ and
$k_{0}= 10^{7}m^{-1}$ for  $k_{0}r_{m}=0.05, 0.07,$ and $0.1$. Near resonance, the transport time depends weakly on the detuning. But, away from
it, $\tau_{tr}$ depends on the detuning and thus on
 frequency.}
\label{fig5}
\end{figure}

\section{VI. Average Self-Energy}
So far, we have used the effective approach introduced in subsection
III-C, where we have considered the case of a scalar wave being scattered by a pair of two-level atoms. 
In this simple approach, the propagator of a scalar wave (\ref{eq23}) has been calculated and the self-energy  (\ref{eq33}) has been obtained by averaging  (\ref{eq23}) over the distance between the  two atoms in a pair.
This effective approach leads to simple expressions for the elastic mean free path  (\ref{eq36}) and the group velocity  (\ref{eq43}) of the wave.
In this section we calculate these quantities for a given $\Delta m$ transition and
$k_{0}r \ll 1$, while taking into account the vectorial  nature of the wave. To this purpose, we average the propagator (\ref{eq22}) over the random orientations of the pairs of atoms (with respect to the wave vector of the incident photon) as well as over the distance between the two atoms in a pair.
Therefore, the average self-energy is now given by
  \be\Sigma_{1}' =  {6 \pi n_i\over k_0}\frac{1}{4\pi r_m}\int{\hbar \Gamma
G^+\textbf{(r)}d\textbf{r}} \label{eq51}\ee where the averaging is
over the inter-atomic axis $\bf r$ (both  over magnitude and
orientations). The evaluation of (\ref{eq51}) for a $\Delta m=0$
transition is rather cumbersome and it is presented in Appendix B. By following the procedure
described in the previous section, we obtain the corresponding
elastic mean free path $l_{e}'$ and the group velocity $v_{g}'$.
In Figure 6 the ratio between $l_{0}$ given by (\ref{eq38}) and
$l_{e}'$ is plotted as a function of the reduced detuning $\delta/\Gamma$ for several values of
$k_{0}r_{m}$.

\begin{figure}[ht]
\centerline{ \epsfxsize 6cm \epsffile{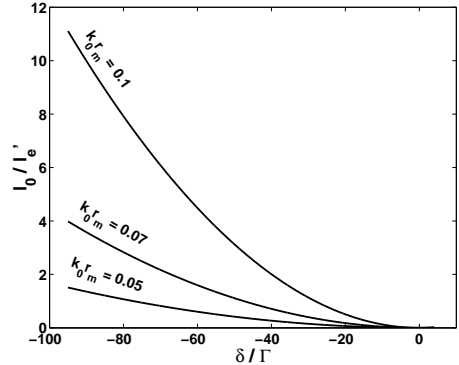} } \caption{\em
Ratio between the elastic mean free paths $l_{0}$ and  $l_{e}'$ as a function of the reduced detuning $\delta/\Gamma$ for $k_{0}r_{m}=0.05, 0.07,$ and $0.1$. At resonance $l_{e}'$ is 
larger than $l_{0}$, but away from resonance it becomes
smaller.} \label{fig6}
\end{figure}

As in the effective approach, at resonance $l_{e}'$ is found to be
larger than $l_{0}$, but away from resonance it becomes
smaller. 

In Figure 7 the group velocity $v_{g}'$ is plotted as a function  of
the reduced detuning $\delta/\Gamma$ for ${n_i / n_c}=10^5$ and $k_{0}r_{m}=0.1$.

\begin{figure}[ht]
\centerline{ \epsfxsize 6cm \epsffile{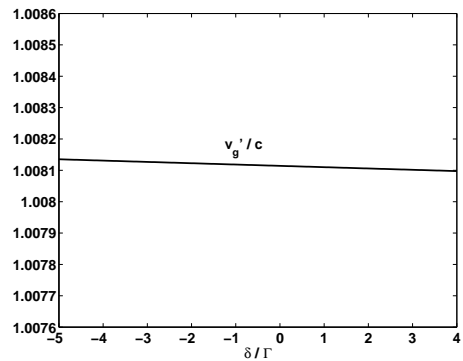} } \caption{\em 
Group velocity $v_{g}'$ as a function of the reduced detuning $\delta/\Gamma$ for ${n_i / n_c}=10^5$
and $k_{0}r_{m}=0.1$. Around resonance, the group velocity $v_{g}'$ is finite and
positive and it  is close to $c$.} \label{fig7}
\end{figure}
Around resonance, the group velocity $v_{g}'$ is finite and
positive, as in the scalar case, but much larger as compared to (\ref{eq43}) and it is close to $c$.
Thus, we may conclude that in both approaches the superradiant effect leads to a finite and positive group velocity, unlike the one obtained for light interaction with independent atoms.   However, the group velocity of a scalar wave is much smaller compared to the one of a photon.

\section{VII. Discussion}
In this section we
compare our analysis to other approaches
\cite{bart3,morice} where resonant multiple scattering of light
has been considered. There, using a multiple scattering expansion
for the calculation of the self-energy up to second order in
$n_{i}u_{0}^{2}$, a correction to the elastic mean free path and
to the refractive index has been obtained. In the latter approach, no
distinction has been made between the external photon that
performs multiple scattering on all atoms and  virtual photons
exchanged between two atoms in a superradiant state leading to the
average interaction potential $V_e$. This distinction needs to be
made for dilute enough atomic gases since in that case the average
distance $n_i ^{-1/3}$ between atoms is large. Moreover, in this
case, the dipole-dipole interaction induced by the external photon
depends on the detuning, a situation that corresponds to the case
of intense radiation presented in \cite{thiru} but not to the
current experiments made on cold atomic clouds \cite{vaujour}.

\section{VIII. Conclusions}
We have considered multiple scattering of a photon by pairs of
atoms that are in a superradiant state. On average over disorder
configurations, an attractive interaction potential builds up
between close enough atoms, that decays like $1/r$. The
contribution of superradiant pairs, resulting from this potential,
to  scattering properties is  significantly different  from that
of independent atoms.  
This shows up in the behaviors of the group velocity, the elastic mean free path and the  diffusion coefficient which are different from their values obtained for independent atoms.
We have considered the case of a scalar wave and have shown that it allows to define an effective long range and attractive potential for pairs of atoms in a superradiant state. Then, we have studied the case of a vector wave and have shown that the results obtained in the scalar case remain qualitatively valid. We have considered a simplified model where only pairs of atoms have been taken into account.  A more realistic model should include higher order terms that account for cooperative effects between more than two atoms \cite{up}.  The purpose of the current model is to show that already for a dilute gas in the weak disorder limit, cooperative effects modify significantly the transport properties of light.

\section{Acknowledgements}
This research is supported in part by the Israel Academy of
Sciences and by the Fund for Promotion of Research at the
Technion. We would like to thank  Jean-Noel Fuchs for constructive remarks.

\setcounter{equation}{0}
\renewcommand{\theequation}{A\arabic{equation}}
\section{Appendix A}
In this Appendix,  we establish expressions (\ref{eq36}) and (\ref{eq43}) for the elastic mean free path  and the group velocity.  At resonance, simple expressions for the elastic mean free path (\ref{eq39}) and the group velocity (\ref{eq47}) are obtained by a pertubative expansion with respect to the small parameter $k_0r_m$.
\subsection{1. Elastic mean free path}
The elastic mean free path  is given  by (\ref{eq36}) in terms of the function $f_1$ defined in (\ref{eq37}). The integral in (\ref{eq37}) is easily carried out analytically and it leads to
\be
\frac{1}{l_{e}(\delta)}=\frac{6\pi
n_{i}}{k_{0}^{2}}\frac{1}{aC_1}\left(\frac{b}{2a}A_1+\frac{a-2}{2a}B_1+C_1\right)\ee
\label{A1} where \be \begin{array}{cc} a=1+(\delta / \Gamma)^{2}
& b=\delta/\Gamma\end{array}\ee
 \be
A_1=\ln \left( \frac{\frac{1}{4}x_{m}^{2}}{a+bx_{m}+\frac{1}{4}x_{m}^{2}}\right)\ee
\be B_1=\frac{\pi}{2}-\tan^{-1}(b+\frac{1}{2}x_{m})\label{A1a}\ee and \be
C_1=\frac{1}{x_{m}}=k_{0}r_{m}\ll1.\ee
 At resonance $(\delta =0)$ we
have $a=1$, $b=0$ and by expanding (\ref{A1a})  with respect to $k_0r_m$ we obtain \be
B_1\simeq\frac{2}{x_{m}}\left(1-\frac{4}{3x_{m}^{2}}\right).\ee Thus,
\be l_{e}(0)=\frac{k_{0}^{2}}{8\pi
n_{i}}\frac{1}{(k_{0}r_{m})^2}\ee as given in  (\ref{eq39}).
\subsection{2. Group velocity}
The group velocity  is given  by (\ref{eq43}) in terms of the function $f_2$ defined in (\ref{eq45}). The integral in (\ref{eq45}) is easily carried out analytically and it yields \be
\frac{c}{v_{g}(\delta)}=1-\frac{n_{i}}{n_{c}}\frac{F_1}{a^{2}C_1}\ee
where

\be
F_1=b(\frac{1}{a}-\frac{1}{4})A_1+\frac{a-2}{4}A_1'+(\frac{3}{2}-\frac{2}{a})B_1-bB_1'+(1-\frac{a}{2})C_1\ee
\be
A_1'=-\frac{b+\frac{1}{2}x_{m}}{a+bx_{m}+\frac{1}{4}x_{m}^{2}}\label{A1b} \ee and \be B_1'=-\frac{\frac{1}{2}}{1+(b+\frac{1}{2}x_{m})^{2}}.\ee
 At resonance $(\delta =0)$ we
have $a=1$, $b=0$ and by expanding (\ref{A1b}) and (\ref{A1a})  with respect to $k_0r_m$ we
obtain \be
A_1'\simeq\frac{2}{x_{m}}\left(\frac{4}{x_{m}^{2}}-1\right)\ee and \be
B_1\simeq\frac{2}{x_{m}}\left(1-\frac{4}{3x_{m}^{2}}\right).\ee Thus,
\be
\frac{c}{v_{g}(0)}=1+\frac{2}{3}\frac{n_{i}}{n_{c}}(k_{0}r_{m})^{2}
\ee as given in (\ref{eq47}).

\setcounter{equation}{0}
\renewcommand{\theequation}{B\arabic{equation}}
\section{Appendix B}
The aim of this Appendix is to calculate the average self-energy (\ref{eq51}) for a  $\Delta m=0$  transition in the case where $k_{0}r\ll1$.
First, we average  the superradiative propagator (\ref{eq22}) over the orientation of the inter-atomic axis and obtain analytical expressions for its real and imaginary parts.
Then, by averaging over the inter-atomic distance  up to $r_{m}$, we obtain the average self-energy (\ref{eq51}).

For a  $\Delta m=0$  transition and $k_{0}r\ll1$, the superradiative propagator (\ref{eq22}) may be written with the
help of (\ref{eq10}) and (\ref{eq11}) as \be \hbar \Gamma G ^{+} =\left[ 
\frac{\delta}{\Gamma}
+\frac{3}{4}\left(\frac{3\cos^{2}\theta-1}{(k_{0}r)^{3}}+
\frac{\frac{1}{2}(1+\cos^{2}\theta)}{k_{0}r}\right)+i\right]^{-1}\ee
 where the inter-atomic axis is $\textbf{r}=(r,\theta,\varphi)$.
An averaging over the orientations 
\be\hbar \Gamma \langle G^+ \rangle =\frac{1}{4\pi}\int \hbar \Gamma G^+d\cos\theta d\varphi \ee  yields for the imaginary  part, \be \hbar \Gamma
\mbox{Im}\langle G^+ \rangle = -\frac{P+Q}{\beta^{2}}\label{eqB2}\ee
and  for the real part, \be \hbar \Gamma \mbox{Re}\langle G^+ \rangle =
W_-P+W_+Q\label{eqB3}\ee where we have defined \be
P=\frac{1}{8A_2\beta \cos(\frac{\gamma}{2})}{\ln\left(\frac{1+2\beta \cos(\frac{\gamma}{2})+\beta^{2}}{1-2\beta
\cos(\frac{\gamma}{2})+\beta^{2}}\right)}\ee
 \be Q=\frac{1}{4A_2\beta
\sin(\frac{\gamma}{2})}\left(\frac{\pi}{2}+\tan^{-1}\frac{1-\beta^{2}}{2\beta\sin(\frac{\gamma}{2})}\right)\ee
 and \be W\pm=-\sqrt{A_2}(\cos\gamma\mp1).\ee
 The auxiliary parameters are given by \be \begin{array}{cc} \beta=({\frac{C_2}{A_2}})^\frac{1}{4}
& \gamma=\cos^{-1}(-\frac{B_2}{2\sqrt{A_2C_2}})\end{array} \ee where \be
A_2=\frac{9}{16(k_{0}r)^2}\left(\frac{3}{(k_{0}r)^{2}}+\frac{1}{2}\right)^2
\ee \be
B_2=\frac{3}{4k_{0}r}\left(\frac{3}{(k_{0}r)^{2}}+\frac{1}{2}\right)\left[\frac{2\delta}{\Gamma}+\frac{3}{2k_{0}r}
\left(\frac{1}{2}-\frac{1}{(k_{0}r)^2}\right)\right]\ee and
 \be C_2=1+\frac{1}{4}\left[\frac{2\delta}{\Gamma}+\frac{3}{2k_{0}r}
\left(\frac{1}{2}-\frac{1}{(k_{0}r)^2}\right)\right]^{2}.\ee

Finally, we average  (\ref{eqB2}) and (\ref{eqB3}) over the
inter-atomic distance up to $r_{m}$
\be \hbar \Gamma
\mbox{Im}\langle \overline{G^+} \rangle = -\frac{1}{r_m}\int_0^{r_m}dr\frac{P+Q}{\beta^{2}}\ee and
\be \hbar \Gamma \mbox{Re}\langle \overline{G^+ }\rangle =
\frac{1}{r_m}\int_0^{r_m}dr\left(W_-P+W_+Q\right).\ee
The integrals can be evaluated numerically and give the average 
self-energy (\ref{eq51}) since \be\frac{1}{4\pi r_m}\int{\hbar \Gamma
G^+\textbf{(r)}d\textbf{r}}=\hbar \Gamma
\langle \overline{G^+} \rangle.\ee


\begin{thebibliography}{99}
\bibitem{rk} It can be shown using (\ref{eq7}), that the contribution of mixed terms like $ \langle
\phi^{\pm}| G | \phi^{\mp} \rangle$ vanishes.
\bibitem{rk2} Notice, however, that this rewriting of the amplitude $T$ is defined up to a proportionality constant which  accounts for the quantum nature of the problem and the photon polarization.
\end{thebibliography}
    \end{document}